\pdfoutput=1
\documentclass[a4paper,fleqn,usenatbib]{mnras}

\usepackage{newtxtext,newtxmath}

\usepackage[T1]{fontenc}
\usepackage{ae,aecompl}
\usepackage{graphicx}
\usepackage{physics}
\usepackage{amsmath}

\usepackage{graphicx}	
\usepackage{amsmath}	
\usepackage{amssymb}	

\title[Short title, max. 45 characters]{The Final Fate of Supermassive $M \sim 5 \times 10^4 \; M_\odot$ Pop III Stars: Explosion or Collapse?}

\author[C. Nagele et al.]{
Chris Nagele,$^{1}$\thanks{E-mail: christophernagele@gmail.com}
Hideyuki Umeda,$^{1}$
Koh Takahashi $^{2}$
Takashi Yoshida$^{1}$
Kohsuke Sumiyoshi$^{3}$
\\
$^{1}$Department of Astronomy, Graduate School of Science, the University of Tokyo, Tokyo, 113-0033, Japan\\
$^{2}$Max Plank Institute for Gravitational Physics (Albert Einstein Institute), D-14476 Potsdam, Germany\\
$^{3}$National Institute of Technology, Numazu College, Ooka 3600, Numazu, Shizuoka 410-8501, Japan
}

\date{Accepted XXX. Received YYY; in original form ZZZ}

\pubyear{2020}

\begin{document}
\label{firstpage}
\pagerange{\pageref{firstpage}--\pageref{lastpage}}
\maketitle

\begin{abstract}
We investigate the possibility of a supernova in supermassive ($5 \times 10^4 \;M_\odot$) population III stars induced by a general relativistic instability occurring in the helium burning phase. This explosion could occur via rapid helium burning during an early contraction of the isentropic core. Such an explosion would be visible to future telescopes and could disrupt the proposed direct collapse formation channel for early universe supermassive black holes. We simulate first the stellar evolution from hydrogen burning using a 1D stellar evolution code with a post Newtonian approximation; at the point of dynamical collapse, we switch to a 1D (general relativistic) hydrodynamics code with the Misner-Sharpe metric. In opposition to a previous study, we do not find an explosion in the non rotating case, although our model is close to exploding for a similar mass to the explosion in the previous study. When we include slow rotation, we find one exploding model, and we conclude that there likely exist additional exploding models, though they may be rare. 
\end{abstract}




\section{Introduction}

Recent observations of high redshift quasars have confirmed the existence of supermassive black holes (SMBHs) of mass up to $10^{9} \;M_\odot$ as early as redshift seven \citep[e.g.][]{banados2018,wu2015}. The formation mechanism of these objects is an open question and many possibilities have been considered including hyper Eddington accretion of solar mass black holes \citep[e.g.][]{haiman2001}, black hole mergers \citep[e.g.][]{volonteri2010}, dense cluster stellar mergers \citep[e.g.][]{omukai2008}, and primordial black hole mergers \citep[e.g.][]{clesse2015}. We will focus on perhaps the most famous scenario, the direct collapse scenario proposed by \citet{bromm2003}. In particular we will discuss the case where the gas cloud forms a supermassive star (SMS) before collapsing to a black hole (for a comprehensive review, see \citet{woods2019}). 

Studies of supermassive stars in the early universe generally fall into three categories. Rotationally supported supermassive stars \citep{baumgarte1999} are stars which rotate at mass shedding angular velocity. The central temperature never gets high enough for hydrogen burning and the star eventually becomes unstable and collapses into a black hole \citep{shibata2002}. Recently, however, because of the $\Omega \Gamma $ limit, supermassive stars are not thought to be able to rotate near the mass shedding limit \citep{hammerle2018}. Studies of more slowly rotating SMSs are broken up into accreting supermassive stars \citep{hosokawa2012,hosokawa2013,schleicher2013,umeda2016,woods2017,hammerle2018a} and non accreting supermassive stars \citep{fuller1986,montero2012,chen2014}. Further studies investigate the collapse of the SMS to a black hole \citep[e.g.][]{shapiro1979,liu2007}, a process which is a possible source for ultra-long gamma-ray bursts (ULGRBs) \citep{sun2017}, gravitational waves \citep{shibata2016,uchida2017,li2018}, and neutrinos \citep{shi1998,linke2001}. We will focus on the non accreting SMS case, though it is possible that the final result is similar to low accretion rate models with the same final mass.

It is not obvious that stars as massive as $10^{4-6} \;M_\odot$ are stable because they are supported mostly by radiation and can experience general relativistic effects \citep[e.g.][]{shapiro1983}; several works have investigated the stability of supermassive stars. For purely hydrogen stars, \citet{fuller1986} found that stars with zero metallicity were stable below $10^{5} \;M_\odot$. They also found that stars above $5 \times 10^{5} \;M_\odot$ were unstable due to general relativistic effects \citep{chandrasekhar1964} and collapsed into black holes \citep[see also][]{montero2012}.

Stars with mass $10^{4-5} \;M_\odot$ and metallicity $Z=0$ are stable until a large enough $^4$He core is formed. At that point, the star is again unstable due to a general relativistic effect acting on the core and the star collapses into a black hole \citep{chen2014}. However, for one of their models, \citet{chen2014} found an explosion due to rapid $^4$He burning during core contraction. This explosion is interesting because it only occurs in a narrow mass range and so does not pose a significant barrier for SMBH formation via the direct collapse scenario, but does provide a signal for which to search with future telescopes. 

In this paper we investigate this explosion in greater depth for a wide range of masses and for some different initial rotational velocities. We first simulate the evolution of the star, and then switch to a relativistic hydrodynamics code to accurately model the collapse or explosion. We find that according to this 1D treatment, no explosions occur, though rotation changes the picture.

This paper is organized as follows. In Sec. \ref{methods} we outline the details of the two codes and changes we have made to them for this paper. In Sec. \ref{results} we gives examples of both collapse and explosion and discuss the situations in which each would occur. Finally in Sec. \ref{conclusion} we discuss the results.

\begin{figure}

    \includegraphics[width=\columnwidth]{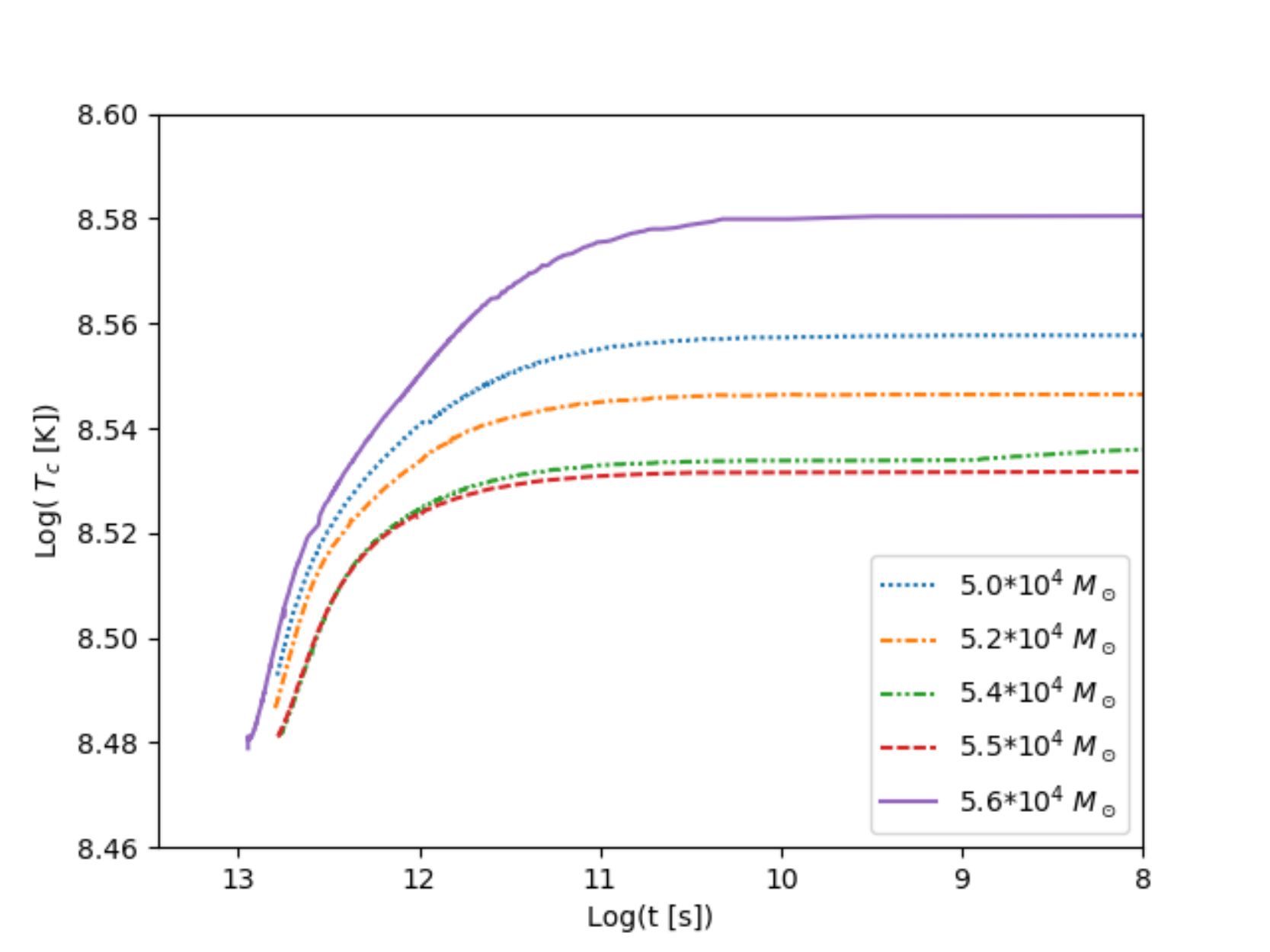}
    \includegraphics[width=\columnwidth]{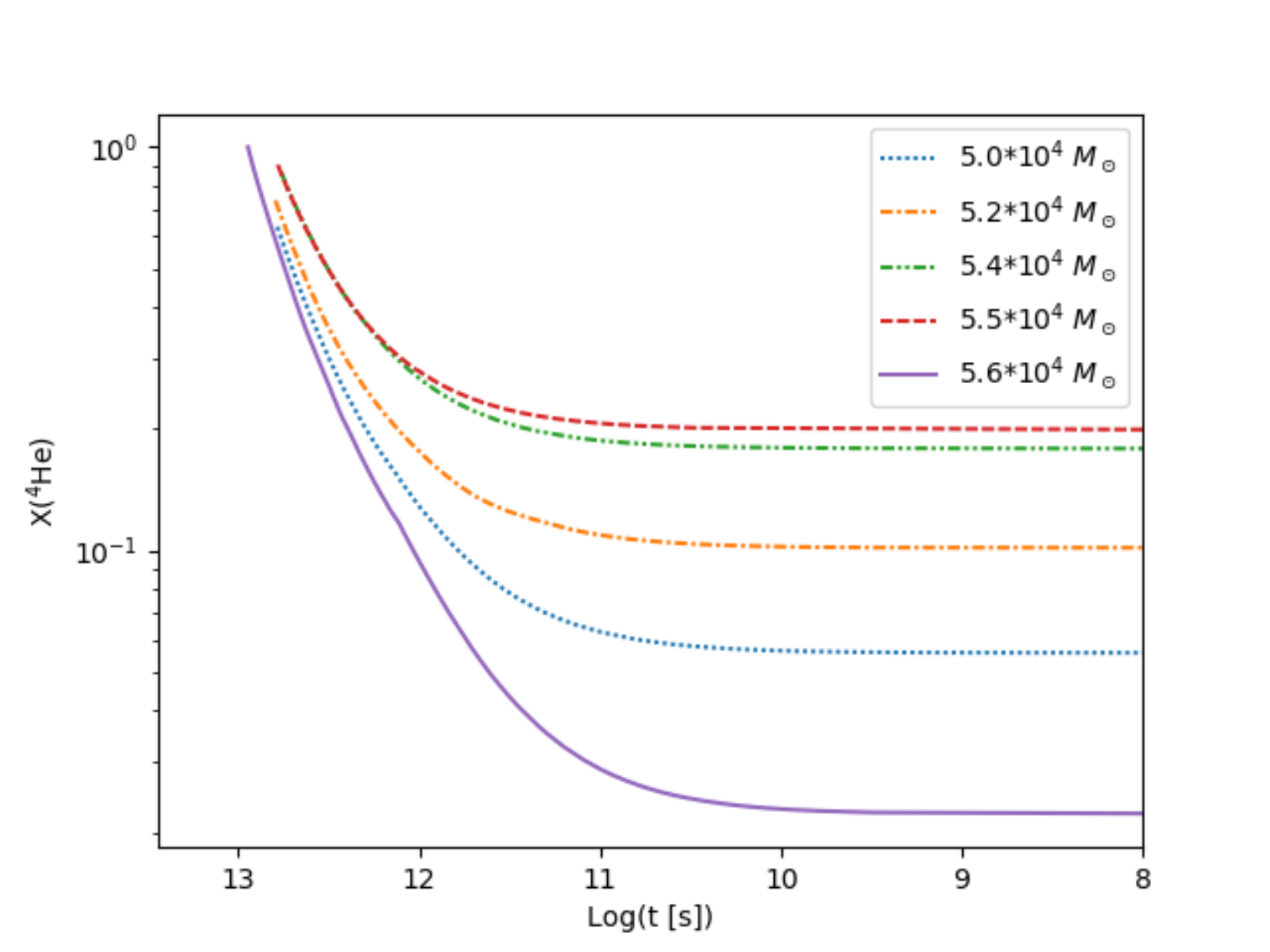}

    \caption{Central temperature (top panel) and central helium mass fraction (bottom panel) plotted versus time until collapse for supermassive population III stars with different masses. These figures begin at the time when $\rho_{\rm c}=\rho_{\rm crit}$ (Eq. \ref{eq:rhocrit}). }
    \label{fig:Logt}
\end{figure}

\begin{figure}

    \includegraphics[width=\columnwidth]{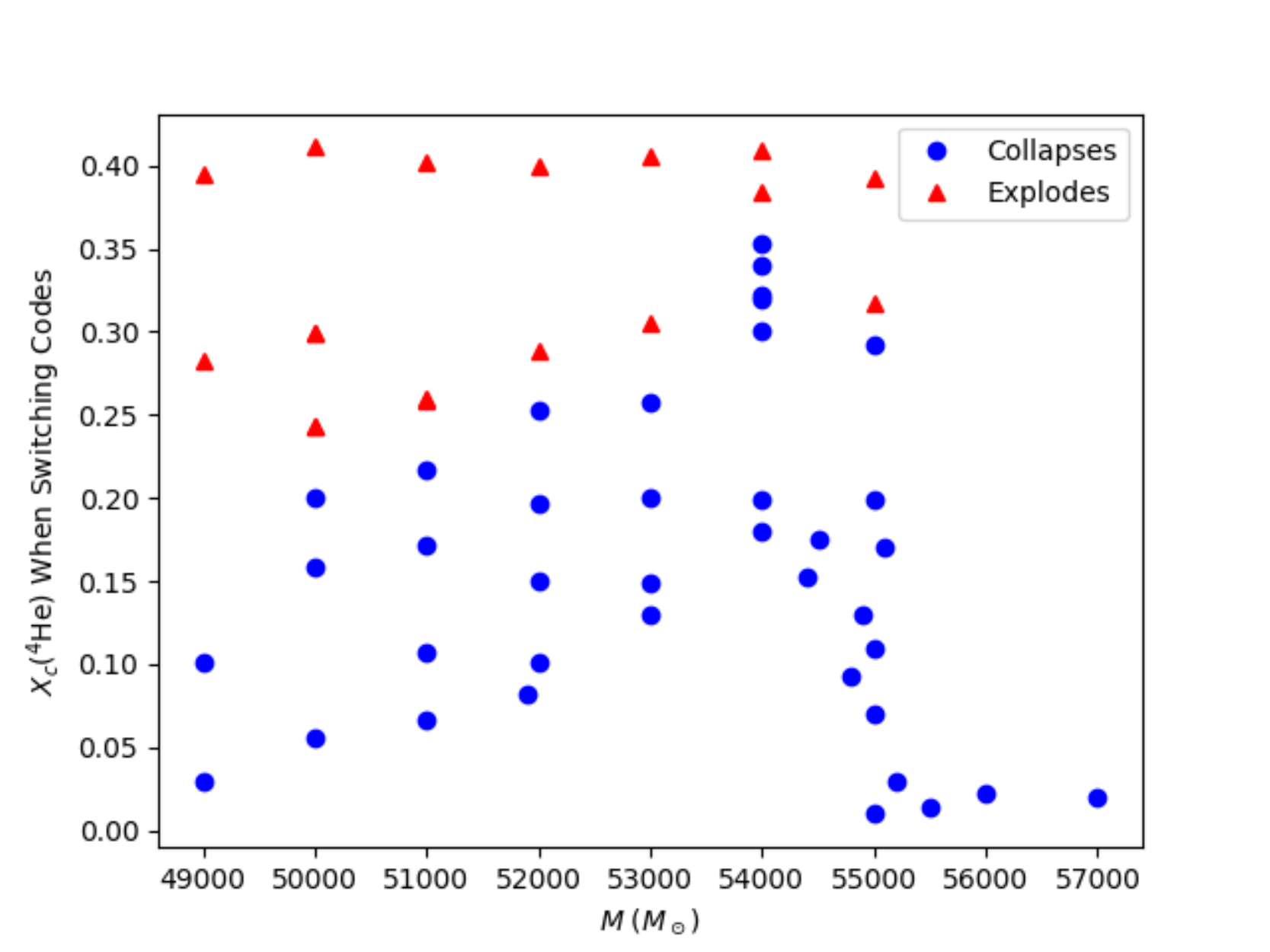}

    \caption{The results of test calculations of non rotating models in HYDnuc showing the outcome as a function of the helium mass fraction when we switch between codes. }
    \label{fig:mass}
\end{figure}

\section{Methods}
\label{methods}
In this paper, we use two codes; one for the stellar evolution calculation and one to simulate the dynamical collapse. 

\subsection{Stellar Evolution}
\label{HOSHI}
A 1D stellar evolution code, HOngo Stellar Hydrodynamics Investigator (HOSHI) is used to evolve supermassive population III stars from hydrogen burning to the onset of gravitational instability \citep{takahashi2016,takahashi2018,takahashi2019,yoshida2019}. In this work, the code uses a 49 isotope nuclear network and does not include mass loss. We assume the direct collapse scenario as the formation channel, and assume no accretion. 

In this paper, $M_{\rm r}$ refers to the mass inside a given radius $r$, $\rho$ is the density, $\mu$ is the mean molecular mass, $s$ the entropy per baryon, $T$ the temperature, and $X(i)$ the mass fraction of an isotope $i$. Quantities with a subscript c such as $\rho_{\rm c}$ refer to the central value of that quantity.

We set rotation at ZAMS to be $\Omega \approx 10^{-15}$ s$^{-1}$. The code then treats rotation self consistently \citep[e.g.][]{heger2000,heger2005,meynet2000,maeder2004}, but we do not expect $\Omega >10^{-10}$ s$^{-1}$ so this is effectively a non rotating case. We investigate the effects of including slow rotation in Sec. \ref{Rotation}. More detailed descriptions about the progenitor evolution and faster rotating cases are given elsewhere (Umeda et al. 2020 in preparation).

We have also added the first order post Newtonian correction to general relativity in the form of the Tolman-Oppenheimer-Volkoff equation \citep{oppenheimer1939,tolman1939}. 
\begin{equation}
    g_{\rm effective} = \frac{-GM_{\rm r}}{r^2}(1+\frac{P}{\rho_0 c^2})(1+\frac{4 \pi r^3 P}{M_{\rm r}c^2})(1-\frac{2GM_{\rm r}}{rc^2})^{-1}
    \label{TOV}
\end{equation}
\begin{equation}
    \nonumber
    \approx  \frac{-GM_{\rm r}}{r^2} (1+\frac{P}{\rho_0 c^2}+\frac{4 \pi r^3 P}{M_{\rm r}c^2}+\frac{2GM_{\rm r}}{rc^2})
\end{equation}
where the second line is the post Newtonian correction and includes only terms of order $c^{-2}$. Our results are dependent on including this correction. In the Newtonian case we found no evidence for an explosion.

\subsection{Dynamical Collapse}
\label{HYDnuc}
After the HOSHI calculation, we switch to a hydrodynamics code with nuclear reactions (HYDnuc). This code is based on a 1D spherical, Lagrangian, general relativistic, neutrino radiation hydrodynamics code (nuRADHYD) developed by \citet{yamada1997} and modified in \citet{sumiyoshi2005}. Nuclear reaction networks were added in \citet{takahashi2016}. In this work, the code uses the same 49 isotope nuclear network as the stellar evolution code and it has an NSE solver for high temperature, as well as neutrino cooling \citep{itoh1996} which is the same as in the stellar evolution code. Since the HYDnuc code does not include radiative transfer of photons, the envelope will collapse if the model is static. For this reason, we can only switch to the hydrodynamics code once dynamical collapse of the core has begun. Switching earlier than this yields collapse starting in the envelope instead of core collapse. 

We report the results of this code partially in terms of the energies of the star. Specifically, the internal energy is 
\begin{equation}
    E_{\rm th}=\int u \; dM_r,
\end{equation}
where u is the internal energy per unit mass, the gravitational energy is 
\begin{equation}
    E_{\rm grav}=- \int g_{\rm effective} \; r \; dM_r,
\end{equation}
where $g_{\rm effective}$ is the Post Newtonian approximation of the TOV equation (Eq. \ref{TOV}), the kinetic energy is 
\begin{equation}
    E_{\rm kin}=\int \frac{v^2}{2} \; dM_r,
\end{equation}
where $v$ is the radial velocity. The total energy is the sum of these three energies
\begin{equation}
    E_{\rm tot}=E_{\rm th}+E_{\rm grav}+ E_{\rm kin}.
\end{equation}
Furthermore, we include the time integral of the energy generation rate
\begin{equation}
    dE_{\rm nuc}=\int \epsilon_{\rm nuc} \; dM_r
\end{equation}
from the start of the HYDnuc calculation
\begin{equation}
    \int_{t_{init}}^t dE_{\rm nuc} \; dt.
\end{equation}

\subsection{Switching Codes}
\label{switch}

At some point during the HOSHI calculation, the star becomes unstable due to the general relativistic instability \citep{chandrasekhar1964} and it is possible for the entire star to experience dynamical collapse. This collapse induces rapid $^4$He burning in the central region and may cause an explosion \citep{chen2014}. Our results are sensitive to the condition for switching between codes; explosion or non-explosion is mainly determined by the $^4$He mass fraction of the central convective core when we switch to the HYDnuc code. If we switch too early, the $^4$He mass fraction is larger and the model tends to explode.

We will now explain our criterion for the switch between codes; to first approximation, the GR instability coincides with the onset of dynamical collapse. There are a few ways of calculating the condition for the instability, but here we will assume that the star is radiation dominated and that the total entropy per baryon is constant in the isentropic core. Then, the entropy is approximately the radiation entropy:

\begin{equation}
    s=s_{\rm r}+s_{\rm g}\approx s_{\rm r}.
\end{equation}
\citep{shapiro1983}. This means the star has the mass distribution of an $n=3$ polytrope. Then, the critical central density above which the star is unstable is \citep[e.g.][]{fuller1986}
\begin{equation}
    \rho_{\rm crit}=426.4 \times (\frac{0.5}{\mu})^{3}(\frac{M_{ \rm core}}{30000 \;M_\odot})^{-7/2} \rm \;\;g\; cm^{-3}
    \label{eq:rhocrit}
\end{equation}
where $M_{ \rm core}$ is the mass of the isentropic core. 

Using $\rho_{\rm c}>\rho_{\rm crit}$ as a criterion for switching codes, however, does not give a realistic result. This is because up to this point we have neglected the effects of nuclear burning. In reality, strong nuclear burning supports the core for some time before dynamical collapse. In our case, strong helium burning prevents rapid collapse even after the condition $\rho_{\rm c} > \rho_{\rm crit}$ is satisfied. As shown in Fig. \ref{fig:Logt}, after the star satisfies $\rho_{\rm c} > \rho_{\rm crit}$, it takes more than $10^{12}$ seconds until collapse and during this period, there are changes in the physical quantities. Specifically, the central $^4$He mass fraction and the temperature change significantly (Fig. \ref{fig:Logt}). 

However, after $10^9$ seconds until collapse, the $^4$He mass fraction is mostly unchanged. In practice, we switch codes around $10^{6-7}$ seconds before collapse, when the timestep is $\Delta t_{\rm HOSHI} \sim 10^{2-3}$ s, though we expect the results to be the same for switching at any time after $10^9$ seconds until collapse; note that the time step is essentially monotonically decreasing in the flat section of Fig. \ref{fig:Logt}. The Kelvin-Helmholtz time scale at this stage,  $\tau_{\rm KH}$, is on the order of $10^9$ s.  

In summary, we switch codes when $\Delta t_{\rm HOSHI} \ll  \tau_{\rm KH}$ which occurs after $\rho_{\rm c}>\rho_{\rm crit}$ and in between these two times, the composition of the star can change significantly.

\section{Results}
\label{results}

Our results fall into two broad categories: models which collapse directly to a black hole and models which explode due to rapid $^4$He burning. The outcome of the HYDnuc calculation is heavily dependent on the central $^4$He fraction at dynamical collapse (Fig. \ref{fig:mass}). In order to determine the threshold value of $X_{\rm c}(^4$He$)$ for an explosion, we calculated the outcome of HYDnuc using several non rotating models for each mass, most of which had unrealistic values of $X_{\rm c}(^4$He$)$. These were produced by switching codes earlier than our criterion--- that is before the start of dynamical collapse. It can thus be said that these models have artificially high helium mass fractions. Using these artificial models in conjunction with our realistic models, we found that the threshold value for an explosion roughly increases with mass, although the increase is not monotonic due to the chaotic nature of the stellar evolution calculation (Fig. \ref{fig:mass}).

\subsection{Collapse}
\label{collapse}

\begin{figure}
	\includegraphics[width=\columnwidth]{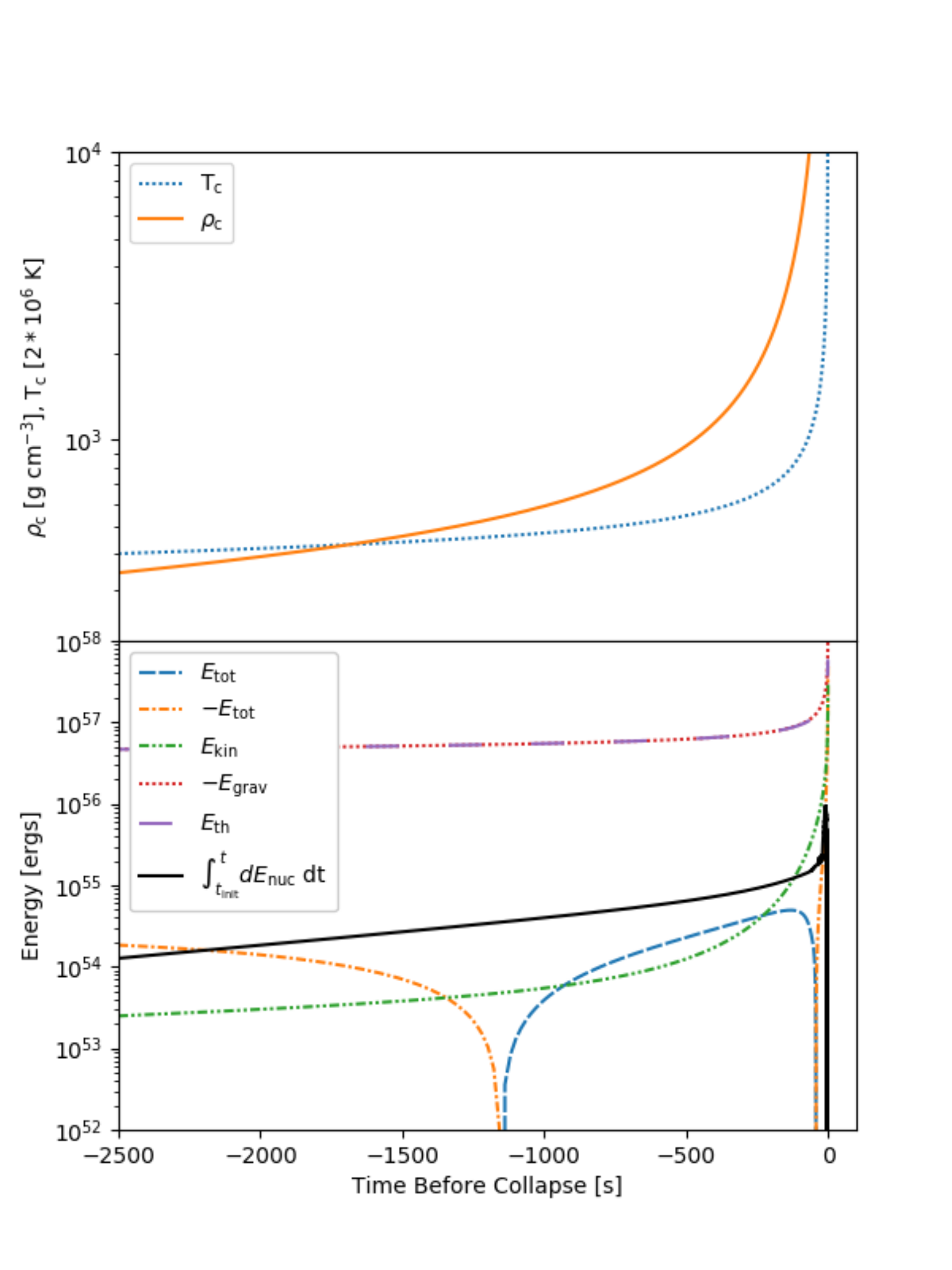}
    \caption{This figure shows an example of collapse to a black hole for a model with $M=5.5 \times 10^4 \; M_\odot$ and $X_{\rm c}(^4$He$)=0.199$ when switching codes. This model is also an example of the non rotating case (see Sec. \ref{HOSHI}) and all of the non rotating models collapsed. The top panel shows the time evolution of central density and temperature while the bottom panel shows the energy quantities from Sec. \ref{HYDnuc}.}
    \label{fig:Ecollapse}
\end{figure}

For most of our models, the star collapses too quickly for the internal energy to become significantly larger than the gravitational energy (Fig. \ref{fig:Ecollapse}). During the helium burning period, central temperature increases and nuclear reactions produce a large amount of energy (see internal energy in Fig. \ref{fig:Ecollapse}), which in some cases is enough to cause the total energy to become positive. However, the total energy does not stay positive for long enough to cause an explosion, and the total energy never becomes significantly larger than the kinetic energy. 42 seconds before the simulation terminates, the total energy once again becomes negative because of a rapid increase in the negative of the gravitational energy, $-E_{\rm grav}$. The calculation terminates due to instability in calculating the NSE around $g_{00}=0.35$ (the time component of the metric) for the central mesh.

We calculate the neutrino cooling luminosity, which peaks at $L_{ \rm \nu,cooling}=1.40 \times 10^{58}$ ergs s$^{-1}$. Because HYDnuc only contains neutrino cooling, this should be viewed as an upper limit for the true luminosity, and we plan to perform a realistic calculation of the neutrino light curve in the future using nuRADHYD. 

It is also possible that the collapse process ejects some heavy elements. In \citet{uchida2017}, which simulated the collapse of a rotating SMS in the helium burning phase, a torus of material was ejected from the collapsing supermassive star; such a torus could contain $^{56}$Ni or other heavy elements if it is ejected in the 10 seconds before black hole formation. However, because the heavy elements are located in the core and because of the slow rotation rate, we do not expect this effect to apply to our models.

\subsection{Explosion}
\label{explosion}

\begin{figure}
	\includegraphics[width=\columnwidth]{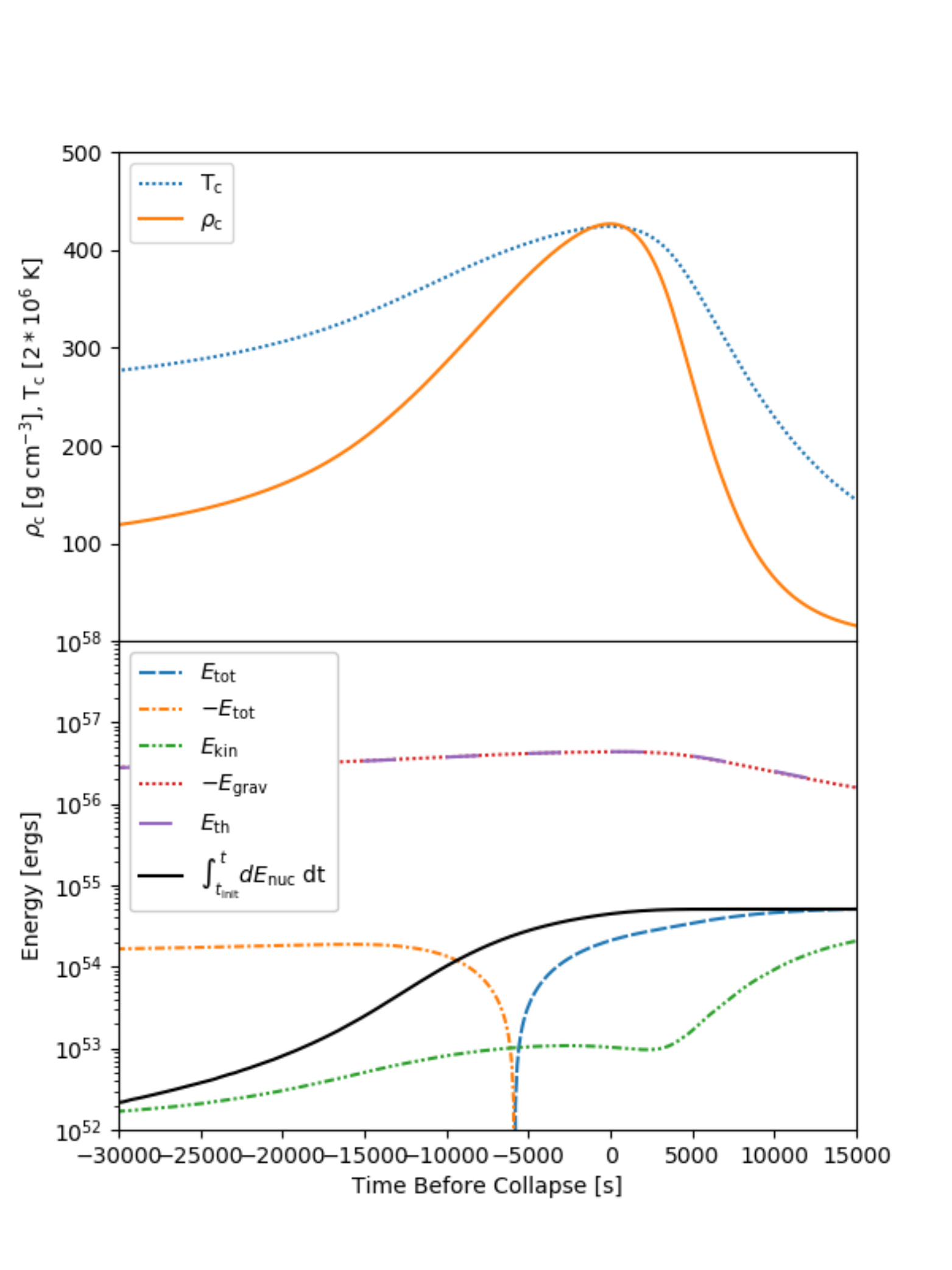}
    \caption{Same as Fig. \ref{fig:Ecollapse}, but for the explosion case. This model has $M=5 \times 10^4 \; M_\odot$ and $X(^4$He$)_{\rm c}=0.190$ when switching codes. It is the only explosion we observed, and it has a slow initial rotation $\Omega = 10^{-10.4}$ s$^{-1}$.
    }
    \label{fig:Eexplosionom}
\end{figure}

We found one realistic exploding model (Fig. \ref{fig:Eexplosionom}). This model had an initial rotation which caused the $^4$He mass fraction at dynamical collapse to be higher than in the non rotating case for this mass (See Sec. \ref{Rotation}). 

 Even though this model has a lower helium mass fraction than the model in the previous section ($X_{\rm c}(^4$He$)=0.190$ compared to 0.199), it is able to explode because it has a lower mass (Fig. \ref{fig:mass}) and because of additional weight from the helium envelope (see Sec. \ref{Rotation}). The initial temperature and density are similar to Fig. \ref{fig:Ecollapse}, but the  kinetic energy is lower (Fig. \ref{fig:Eexplosionom}). This allows the internal energy to overcome the kinetic energy and cause an explosion with explosion energy $E_{\rm tot}=5.45 \times 10^{54}$ ergs. The final outward velocity exceeds the escape velocity in the outer regions of the star, with a maximum value of  $v_{\rm max}=3.11 \; v_{\rm esc}$. However, the inner eighty percent of the star ($M_{\rm r}<40200$ $M_\odot$) does not exceed the escape velocity. Thus the final fate of this object after the explosion is unknown, but it is clear that a large compact object does not immediately form.

\begin{figure}

    \includegraphics[width=\columnwidth]{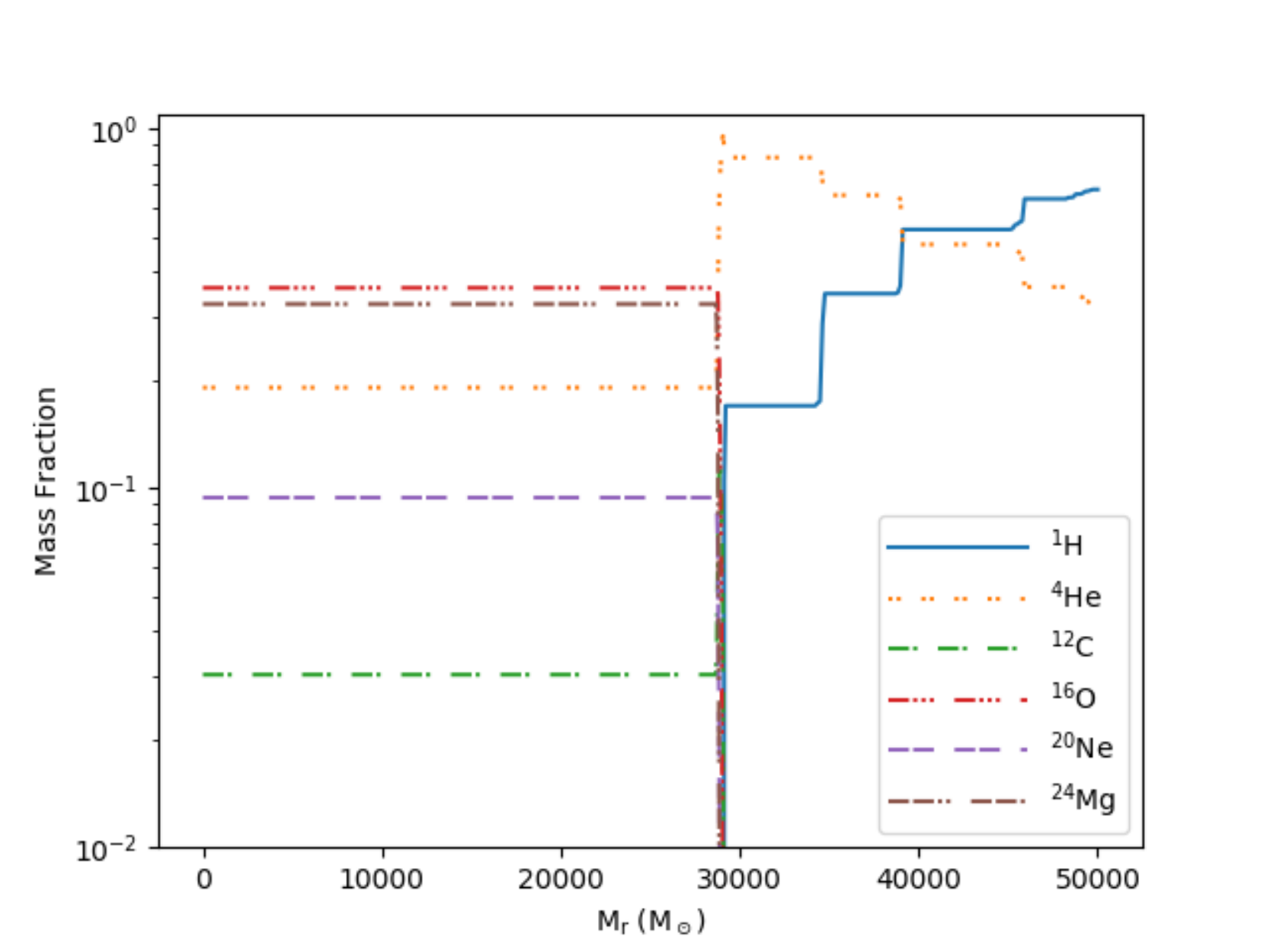}
    \includegraphics[width=\columnwidth]{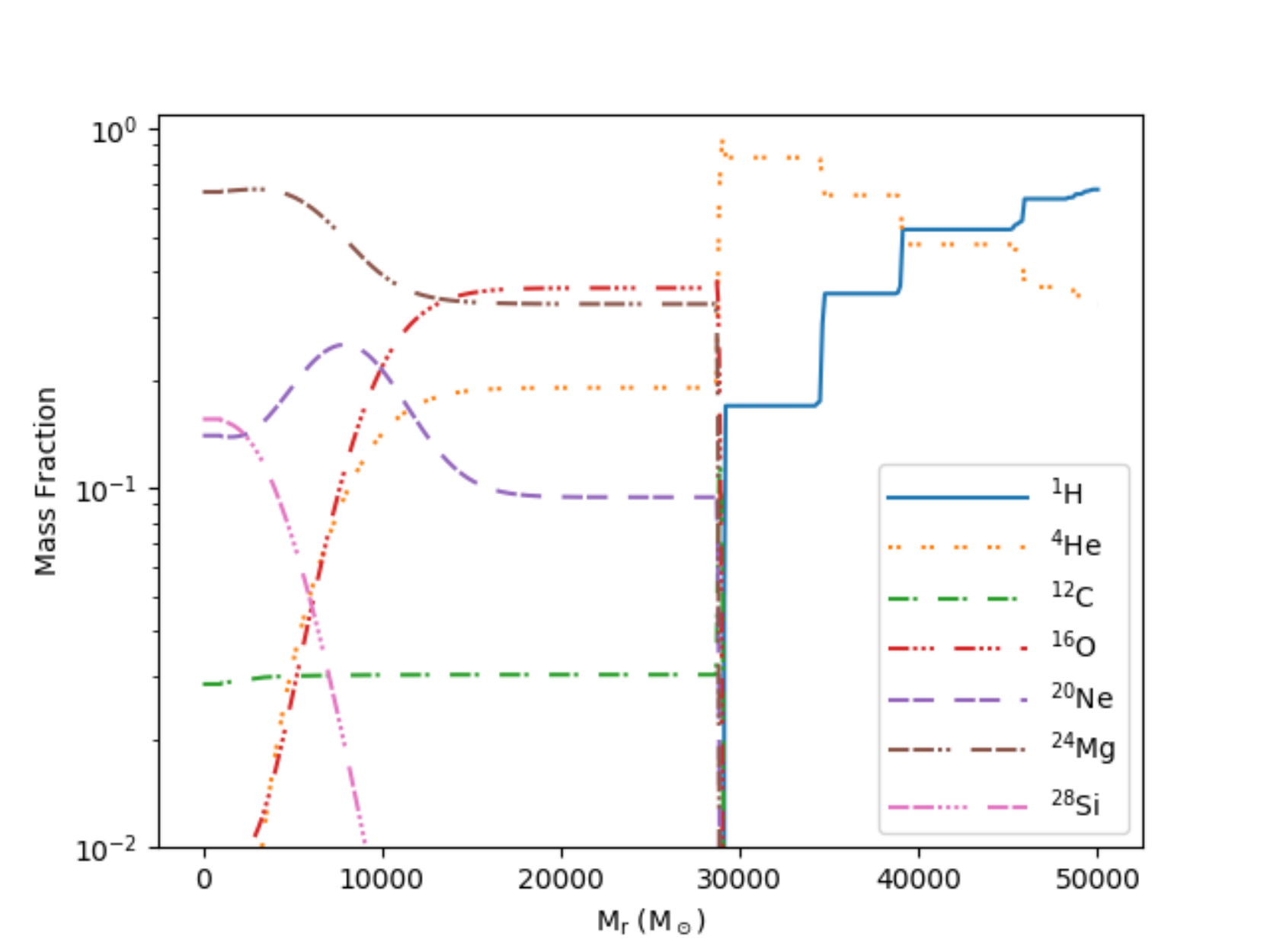}

    \caption{Initial (upper) and final (lower) chemical mass fractions of the HYDnuc calculation for the explosion model (Fig. \ref{fig:Eexplosionom}). }
    \label{fig:chemxpostom}
\end{figure}

The large explosion energy is caused by nuclear burning in a significant proportion of the star during the explosion. In Fig. \ref{fig:Eexplosionom}, helium burning is sustained for 10000 seconds and during this time more than $1500 \;M_\odot$ of $^4$He and $3300 \;M_\odot$ of $^{16}$O is burnt (Table \ref{tab:cxmassom}). The processes responsible for the latter is alpha capture of $^{16}$O, then of $^{20}$Ne, and--- in the inner ten percent of the star--- of $^{24}$Mg (Fig. \ref{fig:chemxpostom}). The explosion energy mostly comes from nuclear burning with $\Delta E=5.10 \times 10^{54}$ ergs.

 In Fig. \ref{fig:chemxpostom}, we can see the initial and final isotope distributions. Since the temperature never gets higher than Log $T_{\rm c}$ [K] $=8.7 $, no elements beyond $^{28}$Si are produced.  The $^4$He-$^{16}$O core is initially convective, as dynamical collapse occurs and central temperature increases, alpha process reactions occur in the center, disrupting the convection.

We should note that the timescale for this explosion is very long ($\sim 10000$ s compared to $\tau_{\rm dynamical} \sim 500$ s for the isentropic core.). In other types of supernovae, such as core collapse or pair instability supernovae, most of the burning takes place on time scales several orders of magnitude shorter. This explosion is caused by the quantity of material burnt, as opposed to reactions producing large amounts of energy quickly.

\subsection{Dependence on Mass}
\label{param}

In the above subsections, we have examined two models, one of which collapses and one of which explodes via rapid helium and oxygen burning. In our simulations, the primary determinant of the fate of the model is the central $^4$He mass fraction when we switch codes. Models with $X_{\rm c}(^4$He$)$ over a certain threshold value explode and those with $X_{\rm c}(^4$He$)$ below this value collapse. We have observed that the threshold value tends to increase with mass (Fig. \ref{fig:mass}), although this is not strictly true due to effects from the envelope which introduce some randomness into our calculations (Sec. \ref{Rotation}). For the mass range considered in this paper, the threshold value is between $0.2<X_{\rm c}(^4$He$)<0.4$. As demonstrated in Fig. \ref{fig:Logt}, toward the end of the calculation in HOSHI, the helium mass fraction in the core decreases due to helium burning, so the later we switch codes, the less helium will be present in the core and thus the star will be more likely to collapse.

\begin{figure}

    \includegraphics[width=\columnwidth]{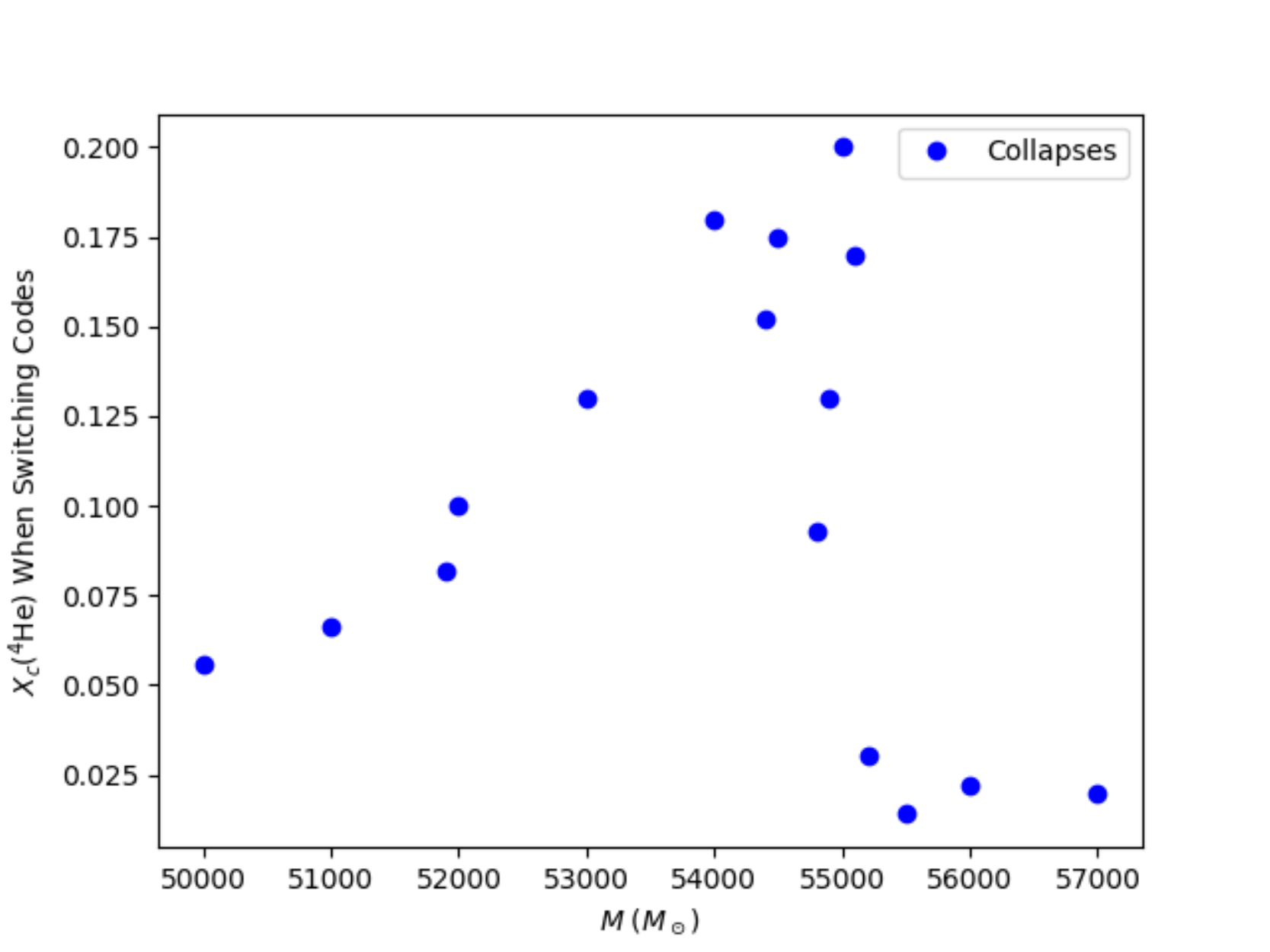}
	
    \caption{This figure shows the dependence of  final $X_{\rm c}(^4$He$)$ from the stellar evolution code on stellar mass for realistic non rotating models.  }
    \label{fig:Heinit}
\end{figure}

We do not find any evidence for an explosion among non rotating stars in the mass range $M=35000-60000 \;M_\odot$. However, we do notice a peak in the central $^4$He mass fraction when changing codes (Fig. \ref{fig:Heinit}) which occurs at $M=55000\; M_\odot$. The location of this peak is near to the exploding model found by \citet{chen2014} at $M=55500\; M_\odot$, though their model had a higher helium mass fraction.

The existence of the peak at $M=55000\; M_\odot$ is interesting and we will now attempt to explain the decrease in helium mass fraction on either side of $55000\; M_\odot$. For $M<55000\; M_\odot$, Eq. \ref{eq:rhocrit} indicates that higher mass means a lower critical density, so that, for increasing mass, the star has a lower central density and--- assuming no large changes in entropy-- central temperature when $\rho_{\rm c}=\rho_{\rm crit}$. Lower central temperature in turn means that less nuclear burning has occurred. Thus as we increase the mass, we also expect to increase $X_{\rm c}(^4$He$)$ at the time of collapse. This can be seen directly in the top panel of Fig. \ref{fig:Logt}, where the initial temperature decreases with increasing mass. Because the rate of helium burning scales with temperature, any increase in temperature is mirrored by a decrease in the central helium mass fraction (the bottom panel of Fig. \ref{fig:Logt}). Thus to have a large helium mass fraction and a higher chance of exploding, a star needs to have a low temperature at the time of collapse. If this $M<55000\; M_\odot$ trend continued to higher mass, we would eventually find models that exploded; however, around $M=55000\; M_\odot$, it is supplanted by a different trend.

For $M>55000\; M_\odot$, when the instability condition is satisfied, $X_{\rm c}(^4$He$) \approx 1$ and $^4$He burning has not yet fully turned on. Without full helium burning, the core cannot support itself against the GR instability (Eq. \ref{eq:rhocrit}) as efficiently and it rapidly contracts to a higher temperature. In the top panel of Fig. \ref{fig:Logt}, the temperature of the $M = 56000\; M_\odot$ model starts lower, but the star is stable for longer than the other models. This is because the sudden jump in temperature after the GR instability releases a large amount of energy which supports the star and allows it to burn helium for significantly longer than the lower mass models (Fig. \ref{fig:Logt}). During this period of stability, helium burning continues and this reduces the central helium mass fraction from $X(^4$He$)\approx 1$ down to $X(^4$He$)\approx .02$

 This effect explains why the dropoff of helium mass fraction at dynamical collapse in Fig. \ref{fig:Heinit} is so sharp for masses greater than $55000\; M_\odot$. Stars in this mass range have almost zero helium available by the time of dynamical collapse, because their helium was burnt during the \textit{stable} period which is--- almost paradoxically--- caused by the jump in temperature after the \textit{instability} equation is satisfied.  

As mentioned above, \citet{chen2014} found an explosion for a model with mass $M=55500\; M_\odot$, which is close to the peak we identify in this paper. For comparison to the model of \citet{chen2014}, we have created an artificial explosion by switching codes much earlier for $55000\; M_\odot$; this mass is the peak discussed above and it comes the closest to exploding of any of our non rotating models. The artificial model is slightly more energetic then the explosion of \citet{chen2014}, but shares the basic characteristics (Table \ref{tab:cxmass}). We are not sure of the cause of the discrepancy in results, but we note that the exploding model in \citet{chen2014} seems to have $X_{\rm c}(^4$He$) \sim 0.4$; such a model would likely explode using our HYDnuc code, but we saw no evidence from the stellar evolution calculation that a model with such a high helium mass fraction is possible.

\subsection{Dependence on Rotation}
\label{Rotation}
\begin{figure}

    \includegraphics[width=\columnwidth]{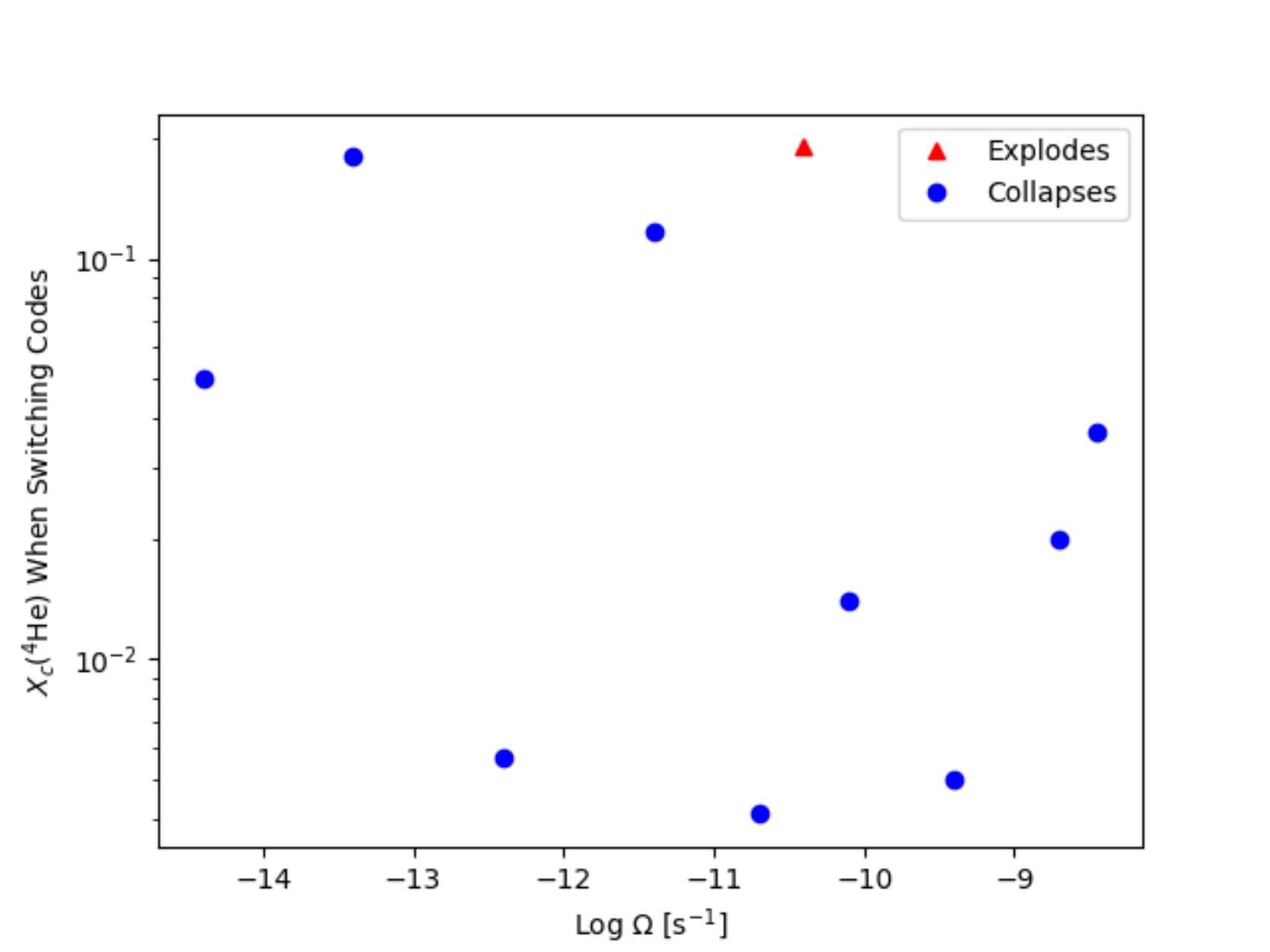}
	
    \caption{The Y axis in this figure is the same as in Fig. \ref{fig:Heinit}. We have plotted this quantity against angular velocity for the $M= 50000 \; M_\odot$ case. The exploding model is the one shown in Fig. \ref{fig:Eexplosionom} and Fig. \ref{fig:chemxpostom}. }
    \label{fig:rot}
\end{figure}

Although rotation cannot be included in the calculation during dynamical collapse, it can be implemented in the stellar evolution calculation. It is reasonable to assume that rotation would stabilize the star and thus prolong the period of $^4$He burning before the onset of collapse. This in turn would greatly reduce the chances of an explosion. However, we find that increased rotation affects the onset of collapse in a non standard way (Fig. \ref{fig:rot}). This figure shows that slow rotation may in some cases lead to larger $X_{\rm c}(^4$He$)$ than in Fig. \ref{fig:Heinit} (the leftmost point in Fig. \ref{fig:rot} is the non rotating case, see Sec. \ref{HOSHI}). The increased helium mass fraction can then result in an explosion, such as the case of $\Omega=10^{-10.4}$ s$^{-1}$ (Fig. \ref{fig:Eexplosionom}).

\begin{figure}

    \includegraphics[width=\columnwidth]{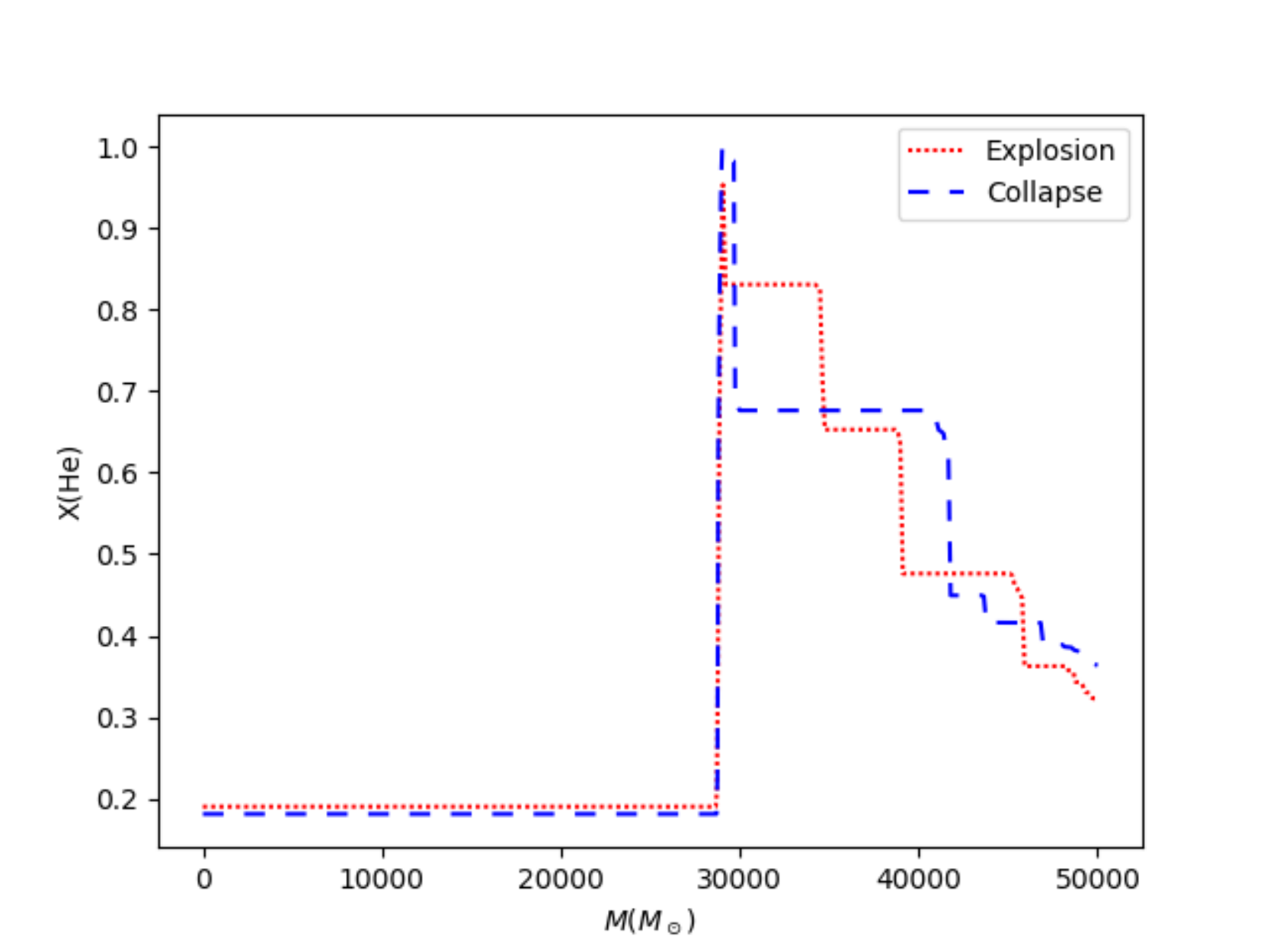}
	
    \caption{The initial helium mass fractions of the models Log $\; \Omega=-13.4$ s$^{-1}$ (collapse) and Log $\; \Omega=-10.4$ s$^{-1}$ (explosion) (Fig. \ref{fig:chemxpostom}). Note that the explosion model has a more massive helium shell outside the core.  }
    \label{fig:cx1e9}
\end{figure}

The behavior in Fig. \ref{fig:rot} is likely due to the effect of convection in the hydrogen and helium envelope; larger convective regions can form a heavy shell just above the core (Fig. \ref{fig:cx1e9}). In Fig. \ref{fig:cx1e9}, the two models shown have similar central helium mass fractions, which is the quantity that we have analyzed up to this point. However, the model which collapses has $M_{\rm final}(^4$He$)=3568 \; M_\odot$ in the 5000 $M_\odot$ outside the core, whereas the explosion model has $M_{\rm final}(^4$He$)=4057 \; M_\odot$. The extra mass in this shell increases the effective weight of the core and can induce collapse earlier than if only the weight of the core is considered. This early collapse can then result in a higher helium fraction and an explosion.

As mentioned above, the size of convective regions in the envelope is stochastic, and this explains the randomness present in Fig. \ref{fig:rot}. Since our investigation of rotation was not very fine grained in the mass or rotation spaces, it is likely that there are other models which explode when slow rotation is introduced, though we suspect they are rare. Finally, the decrease of the threshold $X_{\rm c}(^4$He$)$ value with decreasing mass (Fig. \ref{fig:mass}) suggests that it is easier for an explosion to occur at lower mass, perhaps even below those considered in this paper.

\begin{table*}
	\centering
	\caption{Yields of major isotopes from the explosion of the rotating model in Fig. \ref{fig:Eexplosionom},\ref{fig:chemxpostom} with $M=5 \times 10^4 \; M_\odot$ and $X_{\rm c}(^4$He$)=0.190$ when switching codes. Triple alpha and alpha processes account for most of the burning with the helium oxygen reaction being the most common. Total energy produced is $\Delta E=5.1 \times 10^{54}$ ergs which accounts for most of the explosion energy, $\Delta E/E_{\rm tot}=0.936$. In the below tables, $M_{\rm init}$, is the composition at the beginning of the HYDnuc calculation while $M_{\rm final}$ is the composition at the end of the HYDnuc calculation and $\Delta M = M_{\rm final}-M_{\rm init}$ is their difference. }
	\label{tab:cxmassom}
	\begin{tabular}{l|cccccccr} 
		\hline
		Isotope & $^1$H & $^4$He & $^{12}$C & $^{16}$O & $^{20}$Ne & $^{24}$Mg & $^{28}$Si & Total\\
		\hline
		$M_{\rm init} \; (M_\odot)$& 8780 & 17872 & 897 & 10400 & 2706 & 9341 & 6 & 50002\\
		$M_{\rm final} \; (M_\odot)$& 8775 & 16329 & 891 & 7044 & 3963 & 12185 & 812 & 50002\\
		$\Delta M \; (M_\odot)$& -5 & -1543 & -6 & -3355 &  1257 & 2845 & 806 & 0\\
		\hline
	\end{tabular}
\end{table*}

\begin{table*}
	\centering
	\caption{Yields of major isotopes from the artificial explosion discussed in Sec. \ref{param} with $M=5.5 \times 10^4 \; M_\odot$ and $X_{\rm c}(^4$He$)=0.392$ when switching codes. Explosion energy is $E_{\rm tot}=1.04*10^{55}$ ergs which is slightly more than $\Delta E=9.50 \times 10^{54}$ ergs. In comparison to the real explosion in Table \ref{tab:cxmassom}, the artificial explosion has significantly higher nuclear burning and total energy, but it is challenging to determine if this is from the increased mass or the artificially nature of the model. In addition, this artificial explosion is faster ($\sim 5000$ s) and hotter; the main difference in terms of isotope yields is larger production of $^{28}$Si in the center of the star. }
	\label{tab:cxmass}
	\begin{tabular}{l|cccccccr} 
		\hline
		Isotope & $^1$H & $^4$He & $^{12}$C & $^{16}$O & $^{20}$Ne & $^{24}$Mg & $^{28}$Si & Total\\
		\hline
		$M_{\rm init} \; (M_\odot)$& 9694 & 25937 & 2544 & 12816 & 1792 & 2217 & 1 & 55000\\
		$M_{\rm final} \; (M_\odot)$& 9694 & 23181 & 2264 & 8038 & 3091 & 6316 & 2407 & 55000\\
		$\Delta M \; (M_\odot)$& 0 & -2756 & -280 & -4777 &  1299 & 4099 & 2406 & 0\\
		\hline
	\end{tabular}
\end{table*}

\section{Discussion}
\label{conclusion}
We have simulated non accreting population III supermassive stars in order to determine if they collapse and become seeds for early universe supermassive black holes or if they explode and leave electromagnetic transients which could inform us about conditions in the early universe. 

We find that the results depend highly on the amount of helium present at the onset of dynamical collapse and most of our models have low central helium and collapse to black holes. This value could be dependent on the carbon alpha capture reaction rate which is 1.2 times the rate in \cite{caughlan1988}, and this may explain some of the discrepancy between our results and those of \citet{chen2014}. We plan to investigate this in future work, but suspect that the effect of changing this reaction rate will be minor because the time of collapse is dependent on the balance between nuclear burning and gravity, which may not be affected by this rate.

The largest uncertainties in our calculation likely lie in the amount of nuclear burning which occurs after the GR instability, but before dynamical collapse. The amount of nuclear burning during this time will affect the shape and possibly location of the peak in Fig. (6). Because we have only considered models with slow rotation, we do not expect any deviation from the TOV equations; however, it is possible for multidimensional effects to change the burning rates in the core, as well to change envelope structure. We would note once again that the peak in this paper coincides with the explosion found by \citet{chen2014} to within 500 $M_\odot$, suggesting some agreement between the two codes on this point.

It is possible that other effects could aid in the explosion. Including neutrino transfer will have no effect on the final outcome because the explosion occurs at fairly low temperature. For the model in Fig. \ref{fig:Eexplosionom}, the neutrino luminosity never exceeds $L=10^{48}$ ergs/s. However, multidimensional effects could contribute to an explosion, either through earlier dynamical collapse or providing more helium to the central region during collapse. In addition, a multidimensional treatment would allow for the helium distribution in the core to be less homogeneous, which could lead to a higher central abundance, although it should be noted that \citet{chen2014} did not find any difference between the results of their 1D and 2D calculations. Including radiative transfer during the explosion may also have an effect on the outcome, but due to the high gravitational energy and low temperature during the explosion, we expect this effect to be small.

 We had some concern about the original code not taking convective mixing into account and consequently central $^4$He being exhausted earlier than in reality, which could in theory cause an earlier collapse and affect the outcome. However, after implementing core mixing we found little effect on the condition for an explosion.

Furthermore, we would like to mention numerical scatter for our models, especially given the lack of any trend in Fig. \ref{fig:rot}. We believe that the scatter in Fig. \ref{fig:rot} is definitely due to some aspect of the rotation (c.f. small scatter in Fig. \ref{fig:Heinit}), but should consider that any slight parametric difference could cause the variation in the $^4$He mass fraction.

We have shown that including slow rotation during the stellar evolution calculation has the potential to increase the final helium mass fraction, and to cause an explosion via the action of the convective envelope on the core. Since supermassive stars are thought to be slow or medium rotators \citep{hammerle2018}, this effect will be important. We suspect that more exploding models exist besides the one we have identified, although because we have found only one exploding model for the several hundred realistic models we have tried, it appears to be a quite rare occurrence. 

The explosion or collapse of an individual model has large uncertainties from a number of factors, most notably the envelope structure. However, we have shown that models with masses near 55000 $M_\odot$ come close to exploding, and we can thus say that explosion is a possible outcome for this mass range. Whichever the outcome, these events will likely leave signals observable to future instruments and it is possible that in the coming decades we will be able to directly determine whether or not the direct collapse scenario occurred in the early universe.

\section*{Acknowledgements}

This study was supported in part by the Grant-in-Aid for the Scientific Research of Japan Society for the Promotion of Science (JSPS, Nos. JP 17K05380, JP17H01130, 15K05093, 19K03837), and by Grant-in-Aid for Scientific Research on Innovative areas "Gravitational wave physics and astronomy:Genesis" (17H06357, 17H06365) from the Ministry of Education, Culture, Sports, Science and Technology (MEXT), Japan. For providing high performance computing resources, YITP, Kyoto University is acknowledged. K.S. would also like to acknowledge computing resources at KEK and RCNP Osaka University.




\bibliographystyle{mnras}
\bibliography{bib}




\bsp	
\label{lastpage}
\end{document}